\documentclass[11pt]{article}
\usepackage{moriond,epsfig}

\usepackage{geometry}                
\usepackage{graphicx}
\usepackage{amssymb}
\usepackage{epstopdf}
\usepackage{xspace}

\newcommand{\sqrts} {$\sqrt{s_{NN}}$\xspace}

\def\Journal#1#2#3#4{{#1} {\bf #2}, #3 (#4)}

\def\JPG{\em J. Phys. G}
\def\NPA{{\em Nucl. Phys.} A}
\def\PLB{{\em Phys. Lett.}  B}
\def\PRL{\em Phys. Rev. Lett.}
\def\PRC{{\em Phys. Rev.} C}
\def\ARN{\em Ann. Rev. Nucl. Part. Sci. }

\title{THE RHIC BEAM ENERGY SCAN - STAR'S PERSPECTIVE}
\author{HELEN CAINES - for the STAR Collaboration}
\address{ Yale University,
New Haven, CT, 06520, U.S.A}

\begin{document}
\maketitle
\abstract{The first decade of RHIC running has established the existence of a strongly coupled Quark Gluon Plasma (sQGP), a new state of nuclear matter with partonic degrees of freedom. Theory predicts  how transitions to this sQGP depend on the baryon chemical potential, $\mu_{B}$, and temperature, T. At low $\mu_{B}$ and high T a cross-over transition occurs. At high $\mu_{B}$  and low T the transition is of  first order. Hence, at intermediate values, a critical point should occur. Experimentally we can vary these initial conditions by altering the beam energy. Thus a beam energy scan (BES) will allow us to  explore the  QCD phase diagram close to the QGP-hadron gas boundary  and locate such  key ``landmarks"  as the  critical point. Establishing the existence of  this critical point would be a seminal step forwards for QCD physics.   I discuss below the physics case for a BES, and explain why RHIC and the STAR experiment are ideally designed for such a program.  }

\section{Questions that drive a BES, and the advantages of RHIC and STAR}

The case for a Beam Energy Scan (BES)  rests on four cornerstone questions.
1) Is there a critical point in the QCD phase diagram of nuclear matter and what is its location?
2) Is there evidence of a first order phase transition?
3) At what collision energy does the transition to a sQGP no longer occur?
4) What novel and unexpected physics awaits in the unexplored regions of the QCD phase space?

Before discussing in more detail the physics case for a BES, I will  show why RHIC and the STAR experiment are the ideal accelerator/detector combination for pursing such a program. First, RHIC is a very versatile {\it collider}. This means that the detector acceptance  is independent of collision energy, resulting in many of  the systematic uncertainties canceling, to first order, when data from different  \sqrts are compared.  In addition, the number of particles per unit area at a fixed distance from the collision zone is much lower in a collider setup than in a fixed target one for the same \sqrts, plus the dependence of the occupancy on collision energy is also greatly reduced.  Hence, there are fewer problems related to charge sharing between  hits, and track merging in a collider experiment. All of the above points imply that  we will have excellent control of the systematics of the measurements. 

The STAR experiment with its full azimuthal acceptance for $|\eta| < 1$ and extensive particle identification abilities is uniquely positioned to carry out an energy scan program. With relatively short run periods, high statistics data can be taken that will allow analysis of unprecedented detail  over the energy range planned. Au-Au test runs at \sqrts = 20 and 9 GeV were highly successful both for the RHIC accelerator and STAR. With only a few thousand events taken over the course of a day STAR has been able to report preliminary results ~\cite{kumar,odyniec}.

In this report I highlight a handful of measurements that are among the key results needed to answer the questions posed above. There are, of course, other important  studies that can be performed during a BES at RHIC but they are too numerous to cover in any detail here. Therefore I choose to focus below on  a few analyses to give a taste of the physics that will be revealed by a BES.

\section{Evidence of a Critical Point}

At the critical point extreme long wavelength fluctuations in the susceptibilities of conserved quantities (such as baryon number, charge, and strangeness) are expected to occur~\cite{fluc1,fluc2}. On an event-by-event basis we hope to observe these fluctuations as a function of \sqrts, or $\mu_{B}$/T by measuring the moments of such variables as the   particle ratios (e.g. K/$\pi$ and p/$\pi$), net baryon number, net strangeness etc.  One expects to see a non-monotonic behavior around the critical point. However, the magnitude of these oscillations is hard to predict.  While several of these measurements have been attempted previously at the SPS, STAR's large acceptance allows us to measure such variables with increased sensitivity in each event. Also the ability to  make these measurements in the same detector gives improved control of the systematics as a function of collision energy.

\section{Evidence of an First Order Phase Transition}

The time to reach thermalization, $\tau_{0}$, in RHIC collisions appears to be short, calculations estimate $\tau_{0} \sim$1.2-0.7 fm/c for \sqrts = 5-39 GeV\cite{lifetime}.  Therefore  hydrodynamical models may  appropriately be used  to reveal information about the space-time evolution of the medium. If the lower collision energies have trajectories that cross through a 1$^{st}$-order phase transition, a significant softening of the equation of state is expected.  

 In peripheral events the overlap region of the colliding nuclei is "almond" shaped. Hydrodynamical models predict that the high pressure gradients resulting from this initial spatial anisotropy within the collision zone produce  final state momentum anisotropies, i.e. the medium flows. This generator of flow is self-quenching since the initial driving spatial anisotropy vanishes. A fourier expansion of the  angular distribution of the particles predicts that
$\frac{dN}{d\phi} \propto 1 + 2v_{n}cos[2(\phi-\Psi_{RP})]$ where v$_{1}$ is called directed flow and v$_{2}$ is elliptic flow.  Directed flow is generated during the nuclear passage time, T$_{pass}$, and can therefore probe the onset of  bulk collective dynamics as long as the passage time, T$_{pass} > \tau_{0}$.  T$_{pass}$ can be estimated as 2R/$\gamma$, which then varies from $\sim$5.6-0.35 fm/c for the energies we are considering ({\it i.e.} T$_{pass} >  \tau_{0}$). The rapidity dependence of the direct flow  is of interest since it is  predicted to "wiggle" at mid-rapidity  when passing through a first order phase transition~\cite{v1a,v1b,v1c,v1d,v1e}.

\section{Evidence of the "Turn-off" of sQGP Signatures}
\begin{figure}[htb]
\begin{minipage}{0.4\textwidth}
\begin{center}
\includegraphics[width=\textwidth]{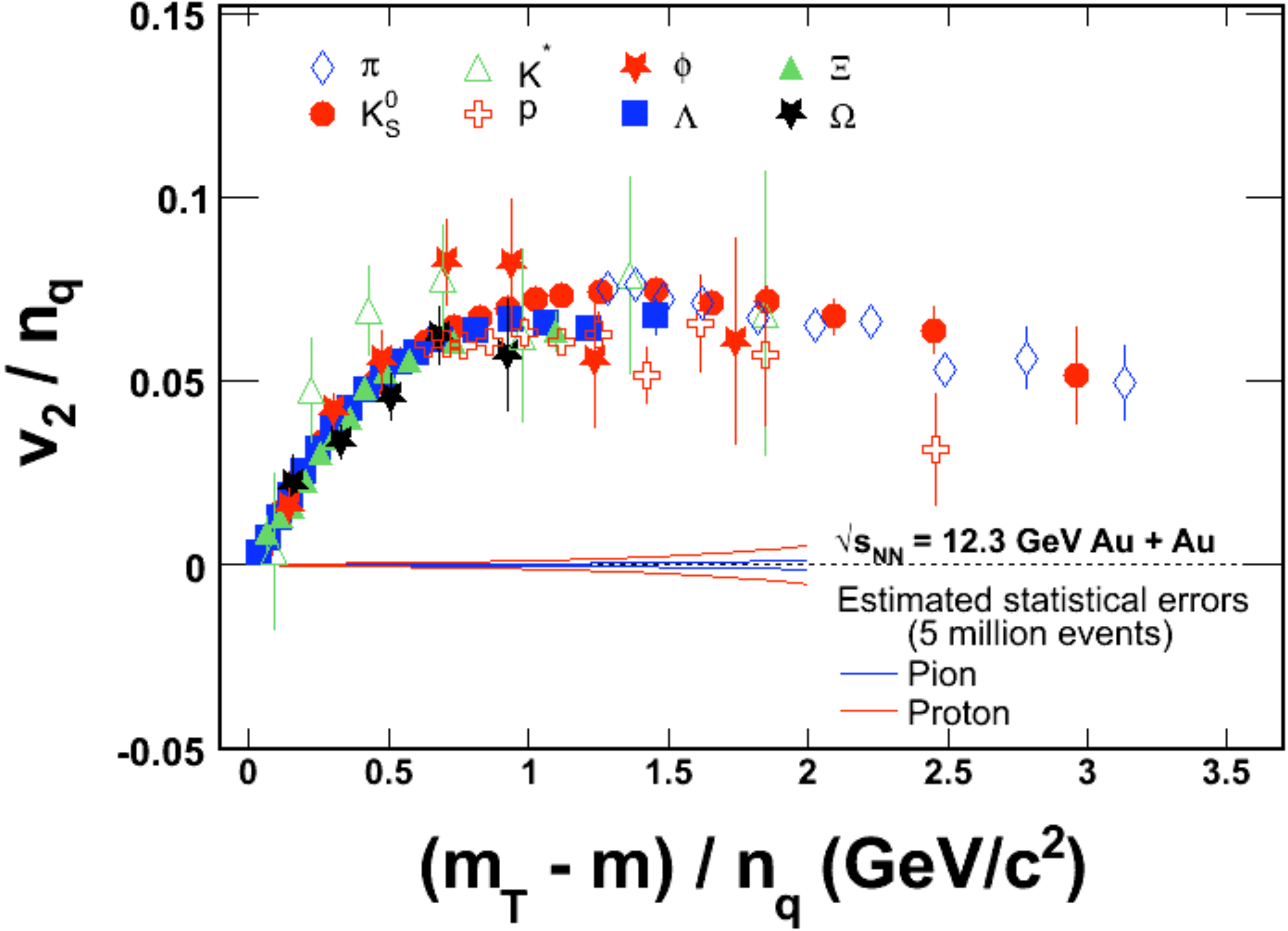}
\end{center}
\caption{Identified particle v$_{2}$ per constituent quark  as a function of m$_{T}$-m$_{0}$ per constituent quark  for Au-Au collisions at \sqrts= 200 GeV. Also shown is an estimation of the statistical error for identified proton and $\pi$ v$_{2}$ for Au-Au collisions at \sqrts= 12.3 GeV. 5M events are assumed and a centrality cut of 0-43.5$\%$.}
\label{Fig:v2KETnq}
\end{minipage}
\hspace{0.5cm}
\begin{minipage}{0.40\textwidth}
\begin{center}
\includegraphics[width=\textwidth]{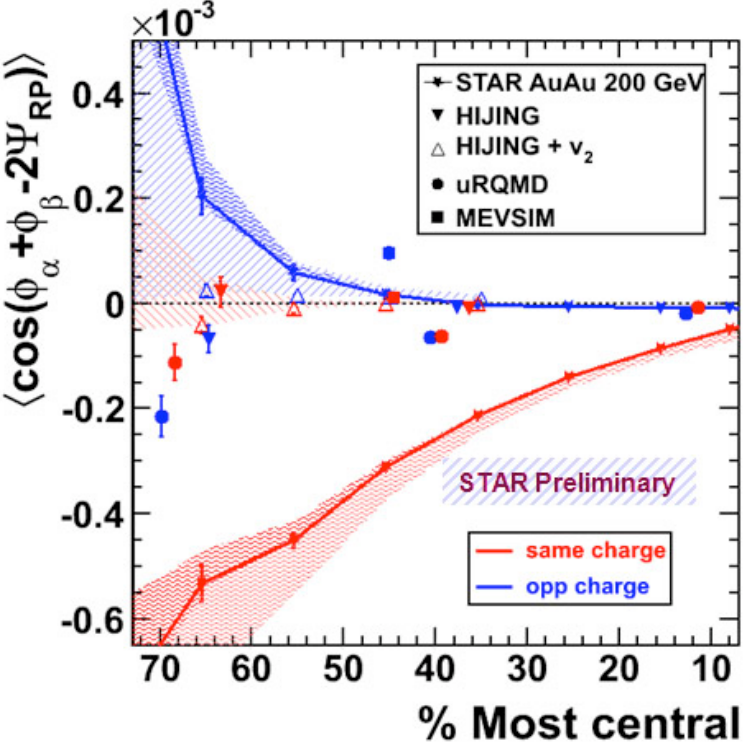}
\end{center}
\caption{The potential local parity violation in strong interactions signal in Au-Au collisions at \sqrts= 200 GeV as a function of centrality.   The shaded bands indicate the systematic uncertainties.}
\label{Fig:ParityViol}
\end{minipage}
\end{figure}

At the lowest beam energies the energy density  could  drop below that required  to produce a sQGP. In these cases we expect to see the disappearance of  signatures  currently believed to  indicate the creation of this new state of matter. One such measurement  is the scaling of elliptic flow with the number of constituent quarks. Elliptic flow, v$_{2}$, is the second harmonic of the fourier expansion of dN/d${\phi}$ described above. As with directed flow, the self quenching of this anisotropy causes v$_{2}$  to be sensitive to the early stages of the medium. The v$_{2}$ of  identified particles as a function of transverse kinetic energy, m$_{T}$-m$_{0}$,  shows  that baryons and mesons follow two different curves.   At intermediate values of m$_{T}$-m$_{0}$ a plateau is reached with baryon v$_{2}>$ meson v$_{2}$.  If however, as shown in Fig.~\ref{Fig:v2KETnq}, one scales both the v$_{2}$ and the transverse kinetic energy,  by the number of constituent quarks in the hadron, {\it all} particles now fall on a common curve ~\cite{Adams:2003am,Adams:2004bi,Adams:2005zg,Adams:2003xp}.  The essential degrees of freedom  at the hadronization seem to be  quarks which have developed a collective elliptic flow during the partonic evolution of the medium.  
  
\section{New Physics}

An exciting recent result  is the suggestion of local  parity violation in strong interactions in  RHIC collisions, predicted to occur if  the QCD vacuum has a non-trivial topological structure - including instantons and sphalerons~\cite{Warringa}. The theoretical motivation for local  parity violation in strong interactions in heavy ion collisions is based on several combined effects.  The  large net charge of the system creates a very intense, localized and short lived magnetic field in peripheral events, due to the large orbital angular momentum perpendicular to the event plane.
 If a sQGP is formed, local strong parity violating  domains can be created leading to an asymmetry in the number of left- and right-handed quarks. The presence of the magnetic field means that there is then a preference, on an event-by-event basis, in the emission direction of like-signed particles along the B-field vector. The angular distribution of charged particles $\frac{dN_{\pm}}{d\phi} \propto 1 + 2a_{\pm}sin[n(\phi-\eta_{R})]+... $, ignoring the v$_{n}$ terms described above, a$_{\pm}$ is the asymmetry due to parity violation. Unfortunately this averages to zero over many events due to the random distribution of the charged domains.
A non-zero parity violation measure  can be obtained by instead measuring $\langle cos(\phi_{\alpha}+\phi_{\beta}-2\Psi_{R})\rangle \approx (v_{1,\alpha}v_{2,\beta}-a_{\alpha}a_{\beta})$. The distribution of this measure as a  function of centrality is shown in Fig.~\ref{Fig:ParityViol}~\cite{parity}. While the results are consistent with  such a parity violation signal it is important to note that the measurement is parity-even and hence could also result from other effects. Several possibilities, such as jets and resonances, have been investigated and to date no background source has been shown capable of producing such a strong signal. Since this signal is dependent on deconfinement, the BES can be used to see if this parity violation measure disappears at the same \sqrts as the other sQGP signals. At higher energies the survival and \sqrts dependence of the measurement can be compared to predictions to further confirm this exciting potential parity violation signal.

\section{Summary}

In summary, there is great potential for important new physics results to emerge from a beam energy scan at RHIC. We hope to pursue this with an broad initial scan in  2010  with collisions at \sqrts=5, 7.7, 11.5, 17.3, 27, and 39 GeV. The lower energies have been selected to further map out the region where the SPS results have been obtained. We hope to return with a finer scan at a later date around  \sqrts energies identified as ``interesting" energies during the first scan.

\end{document}